# TESTS AT 2K OF THE BETA 0.35 SPOKE CRYOMODULE PROTOTYPE WITH THE MTCA.4-BASED LOW LEVEL RF SYSTEM PROTOTYPE FOR THE MYRRHA R&D[*]

C. Joly[†1], S. Berthelot[1], S. Blivet[1], F. Chatelet[1], N. Gandolfo[1], C. Lhomme[1,2], G. Mavilla[1], H. Saugnac[1], G. Olivier[1], M. Pierens[1], J-F. Yaniche[1], CNRS/IN2P3, IJCLab, Orsay, France
F. Bouly[3], O. Bourrion[3], Y. Gomez-Martinez[3], D. Tourres[3], CNRS/IN2P3 LPSC, Grenoble, France
C. Gaudin, J-L. Bolli, I. García-Alfonso, IOxOS Technologies SA, Gland, Switzerland.
P. Della Faille[†], M. Vanderlinden, W. De Cock, SCK CEN, Mol, Belgium
[1]also at Université Paris-Saclay, Orsay, France
[2]also at Accelerators and Cryogenic Systems SAS, Orsay, France
[3]also at Univ. Grenoble Alpes and Grenoble INP, France

*Abstract*
Within the framework of the first phase of MYRRHA (Multi-purpose hYbrid Research Reactor for High-tech Applications) project, called MINERVA, IJCLab was in charge of a fully equipped Spoke cryomodule prototype development, tested at 2K. It integrates two superconducting single spoke cavities, the RF power couplers and the Cold Tuning Systems associated. On the control side, a MTCA.4-based Low Level Radio Frequency (LLRF) system prototype and the Software/EPICS developments has been realized by IJCLab and the SCK CEN in collaboration with the company IOxOS Technologies. The final version of the global system and the results of the tests at 2K will show with some perspectives.

## INTRODUCTION

A cooperation agreement on ADS between the SCK CEN [1] and IN2P3 [2] started in 2017, with the start of MINERVA project [3]. It includes one R&D contract for the design, construction and test of a fully equipped 352.2 MHz Spoke cryomodule for the first LINAC section.

In the framework of this contract, IJCLab designed and developed most of the different parts of this prototype [4-6]. The RF power couplers [7-10] and the C&C board for the Cold Tuning System (CTS) were designed and developed by the LPSC. All the qualifications tests, as well as the assembly of the cryomodule, have been realized by IJCLab team. The tests started at the end of 2022 with four tests, each lasting around a month and covering cryogenic, RF, Fault-tolerance, EPICS supervision, archiving and machine protection aspects.

The Digital LLRF system, MTCA.4-based [11], and more broadly the associated systems such as timing, the MPS, RF amplifiers and EPICS supervision, are being developed in collaboration with LPSC, SCK CEN and IOXOS Technologies. Currently, the LLRF system is used for the preparation and tests of the cryomodule in real operation at 2K without beam.

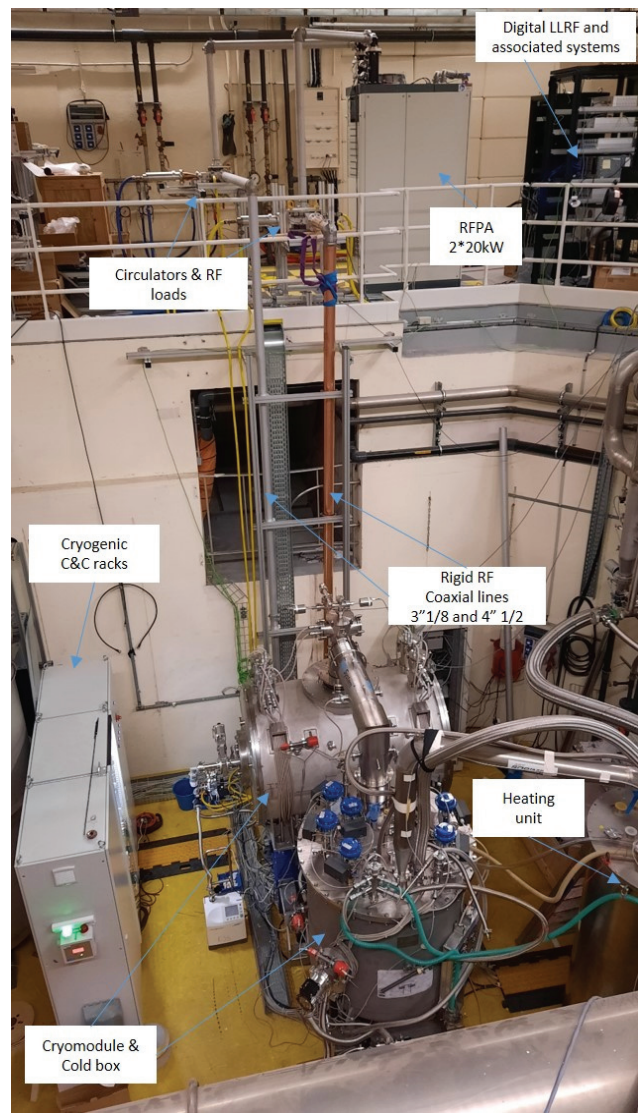

Figure 1: Cryomodule test area at Orsay (SupraTech plateform, IJCLab).

† christophe.joly@ijclab.in2p3.fr / philippe.della.faille@extern.sckcen.be
* Work supported by SCK CEN and IN2P3/CNRS

## CRYOMODULE PROTOTYPE

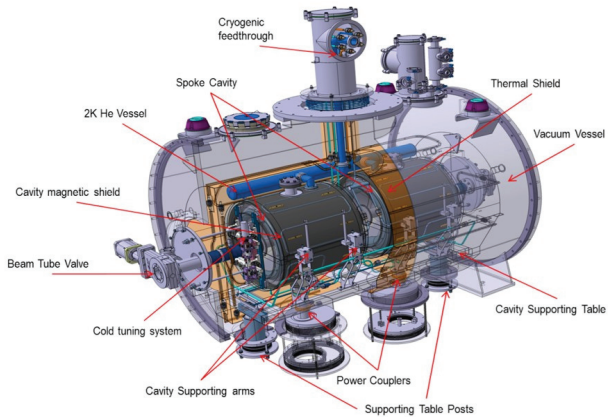

Figure 2: 3D design cryomodule scheme

The conceptual mechanical design of the prototype cryomodule is based on the HB650 PIP-II cryomodule. The cavity string is mounted on a strong back via 'arms' equipped with C-shape blocks for alignment and thermal contraction compensation.

The aim is to simplify assembly and tooling procedures, and to provide predictable cavity string axis alignment to reduce the manufacturing footprint. A short cryomodule design with 2 cavities was chosen to provide greater beam control compliance to meet MINERVA's 'fault tolerance' requirements [12, 13]. The cryogenic active parts (valves and 2K heat exchanger) are placed outside the cryomodule in a dedicated vessel to achieve high modularity for each functional component.

## THE LOW LEVEL RF SYSTEM

As a reminder, apart from the requirements linked with fault-tolerance aspects, the LLRF must also ensure the amplitude and the phase stabilities of the accelerating field, ±0.2% and ±0.1° respectively.

The digital LLRF prototype system is MTCA.4-based and designed around a main board, the IFC1420, and a RF front-end prototype µRTM board called RDC1470 developed in collaboration with IOXOS Technologies. Currently, with the help of this RF Front-End board, the LLRF system has been tested with success on the injector prototype, performing the QWB cavities accelerator field regulation at 176.1MHz, with beam [14]. The validation at 352.2MHz is in progress while operating the two single SPOKE cavities during the Cryomodule tests. For testing them with the cryomodule constraints, some improvements and the addition of specific hardware and VHDL functions were implemented.

Associated to the hardware, the VHDL functions implemented into the FPGA, as shown in Fig. 3 include two types of demodulation, a classic I/Q demodulation (44MHz IF signals sampled at 176.1MHz) or a NON IQ demodulation (9 samples on 2 IF periods). A Proportional Integral (PI) corrector, A Self excited Loop (SEL) and a Generator Driven (GDR) are the three operation modes. Using CORDIC, conversions in Phase and Amplitude allow to implement phase shift calculation, power limitation and Quench detection functions.

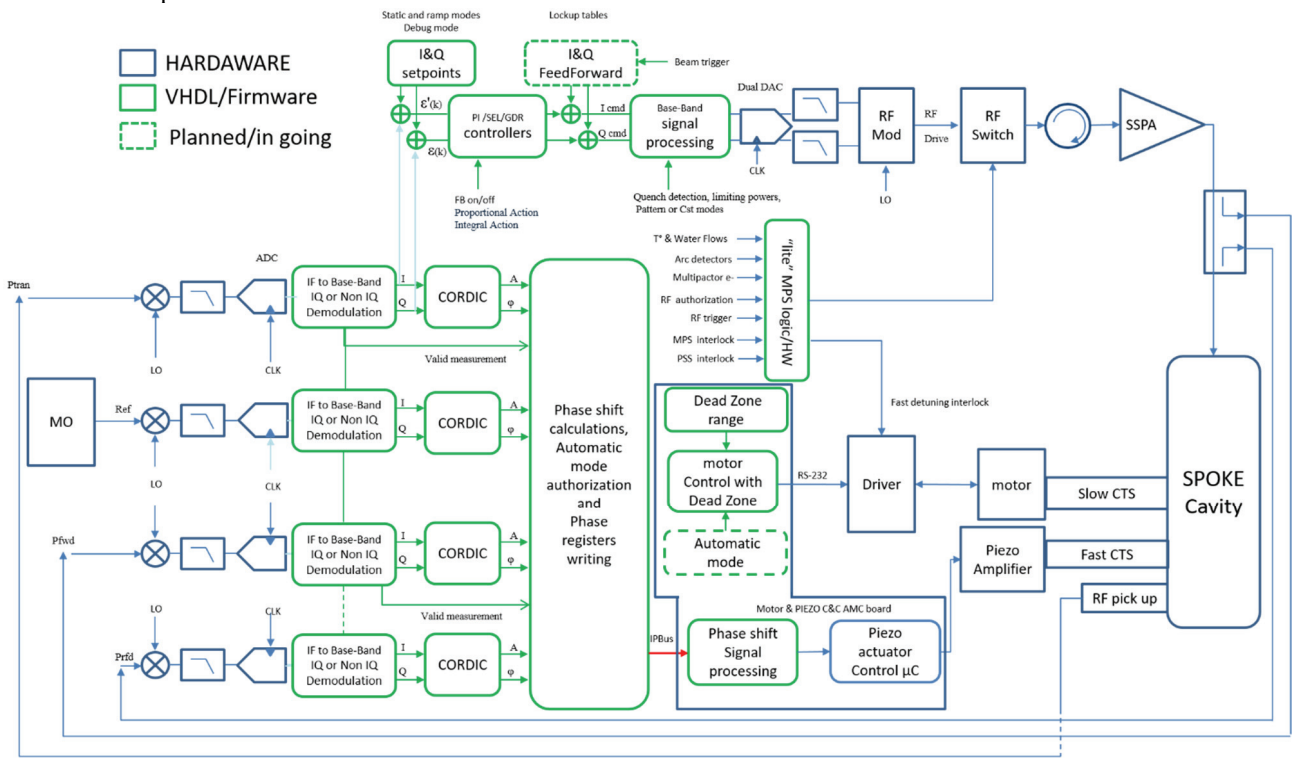

Figure 3: Main hardware and VHDL functions scheme.

# EPICS DEVELOPMENTS

To perform Control and Command (C&C) of the LLRF system, an EPICS Input-Output Controller (IOC) was implemented on the IOxOS IFC 1420 board, using ESS E3 framework [15] with EPICS 7.0.7 [16] but using only the TOSCA libraries. Significant work was done with the IOXOS Technologies team in order to adapt the TOSCA libraries and control the RF front-end board.

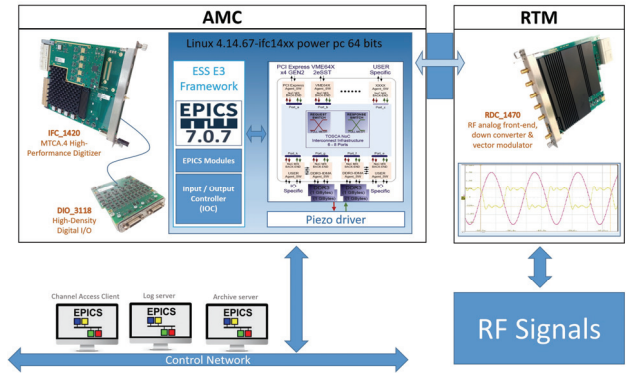

Figure 4: LLRF Control System (CS) architecture.

The embedded OS is a Linux power pc 64 bit build via the open source "Yocto Project" [17].

The ESS E3 framework was chosen mainly for:
- Its management of internal dependencies between the EPICS base and the modules.
- Simplification of its installation in the embedded OS. Easy update and maintenance.
- Ability to run several different versions of EPICS on the same card.

The IOC associated with the FPGA firmware is designed to be multi-functional. It can simply start LLRF regulation on the cavity but also be used as a real oscilloscope which returns data in the form of trigged buffers with a precision of +/- 12 nanoseconds between each measurement. These buffers can thus be archived via EPICS channel access (CA) for retrospective analysis.

The piezo driver installed on the IOxOS OS communicates with the CTS C&C card. We use the C++ Hardware Access Library (uHAL) which provides an API for RMW IPbus reads, writes, and transactions. It is compiled for 64-bit Power PC architecture (see Fig 4). It runs independently from the EPICS layer and enables data exchange with a configurable update rate. The registers exchange can also be configured using an xml file.

The solid-state amplifiers (SSA) [18] driven by the LLRF boards are synchronized via the PTP protocol. The data can also be recovered in the form of trigged buffers sent to the archivers via EPICS.

The SSA cabinet consists of:

- 1-to-n Solid State Module(s) (SSM).
- 8 power supply units (PSU).
- 1 PLC.

The data of SSMs will be directly pulled through Modbus communication by the IOC, whereas the remaining data will be pulled from the amplifier PLC through Ethercat bus communication. The post-processing function is supported by circular buffers as in the LLRF system. A PC with Linux Centos, hosts the EPICS IOCs. These IOCs make the link between EPICS channel access and sub-systems (see Fig 5).

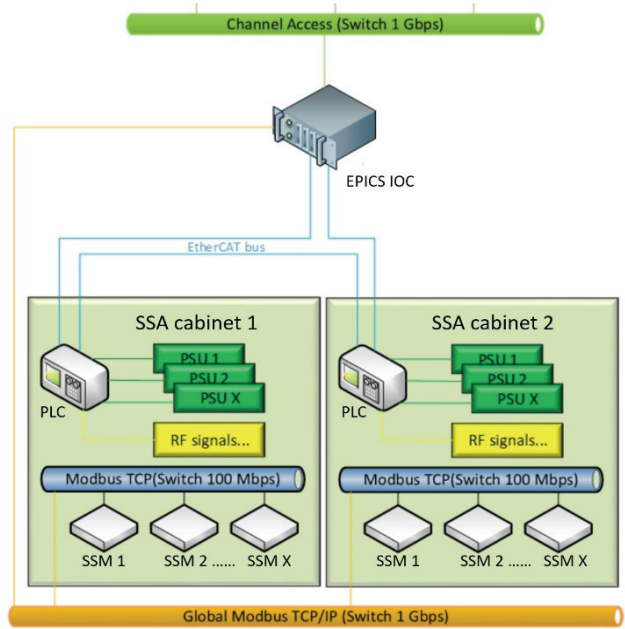

Figure 5: SSA Control System architecture.

The Graphical User Interfaces (GUIs) are developed with Control System Studio 4.6.1.25 (CSS) [19] as illustrated by Fig 6. Configuration files allows to start the LLRF system easily and to manage several similar hardware boards in parallel. It is also possible to change the configuration on-line for specific functions.

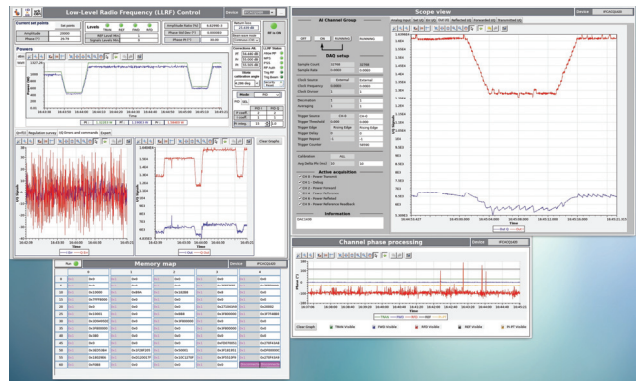

Figure 6: Examples of LLRF control screens

In addition, automatic high-level functions, such as frequency tuning, are implemented using the EPICS sequencer.

## PRELIMINARY RF TEST RESULTS

The first step was the RF power conditioning at 14kW CW of the 2 RF power couplers in the cryomodule at 300K and 4K with the LLRF system in GDR mode and RF Pulsed mode using the MRF timing system. The RF pulses used a width, ranging from 10µs at 16Hz to 99.9 ms at 1Hz. The conditioning was performed out of bandwidth with a first step below the RF power level at which the multipacting current appears, in order to reduce the vacuum level degassing.

The first conditioning at 300K of the RF power coupler without a TiN deposit (CPLR 4) has lasted more than 4 days, (see Fig. 7), allowing to reduce the e- pick up current due to the multipacting effect without canceling it out. For the second RF power coupler (CPLR 2), just one day of conditioning has canceled the e- pick up current.

At 4K, the cryogenic pumping effect made it easier.

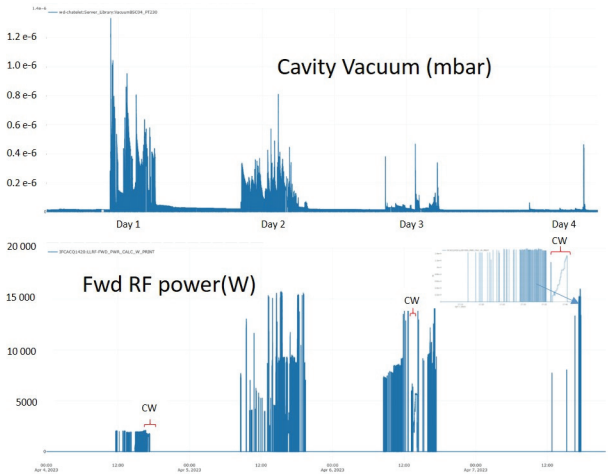

Figure 7: Conditioning at 300K of the CPLR 4.

At 2K, the CTS was used with the LLRF system in GDR mode for tuning the cavity at the accelerator reference frequency, 352.2MHz. The SEL loop has been tested in IQ demodulation mode with success although it is not needed due to the very good frequency stability (Fig. 8) of the spoke cavities prototypes.

The first results are show in Fig 9. The Accelerator field is increased without a frequency regulation, requiring a slow RF power increase using a ramp and a manual tuning adaptation mainly due to the Lorentz forces. In addition, using the mechanical cavity spectrum transformed to a "little music"[†] help us know when to tune the cavity in open loop.

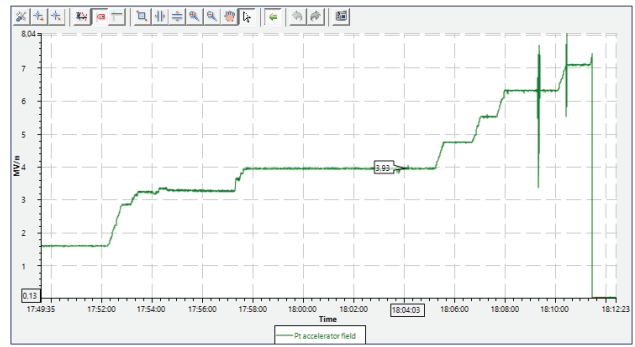

Figure 9: Increase of the Accelerator Field in PI mode using the ramp function with an open loop frequency tuning.

The Phase shift between the forward and transmitted signals, shown in Fig 10, is calculated in the FPGA. It is provided to the supervision and by IPBUS communication to the CTS C&C board. The latter was already validated before the current campaign of tests. We can observe the transition when the stepper motor position changes. The last part of the plot correspond to a quench, and RF is interlocked.

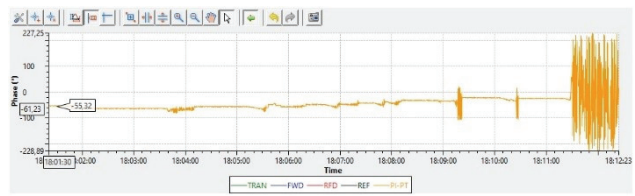

Figure 10 : Phase Shift $P_{fwd}$-$P_{tran}$ showing the cavity detuning, with -55.32° the reference value.

In a next step, the frequency regulation will be tested and some improvements, already planned, will be implemented. Some data exchange will be added for signal level validation to avoid a frequency tuning in case the RF signals are too low or if RF is switched off.

## CONCLUSIONS

Our MTCA.4 LLRF system, already validated in operation at 176.1MHz with beam on the MYRRHA

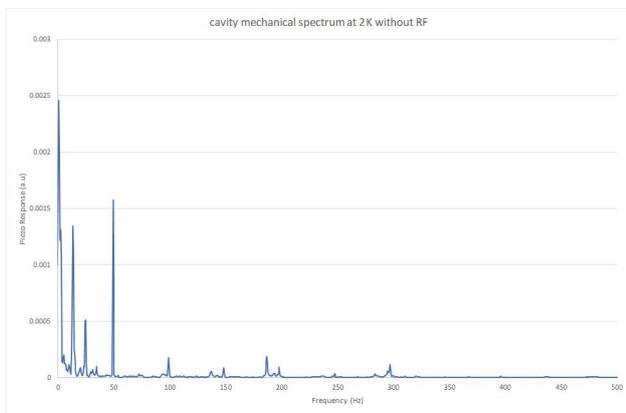

Figure 8: Cavity spectrum without RF at 2K illustrating the frequency stability of the prototype SPOKE cavity.

---

[†] Microphonics are detected by piezo actuators and the signal, in the 1 Hz-10 kHz range, can be listened with a speaker.

injector prototype, is well on the way to being validated at 352.2MHz with superconducting cavities at 2K before the end of this year. Improvements are planned concerning the firmware and software to make it easier to use and integrate more functions. For example, a second version of the RF-Front board prototype, more efficient in terms of frequency Sampling and Local Oscillator frequency range, will be used. Meanwhile R&D is continuing, in particular with developments of a White Rabbit-based Timing system, following the common effort of the community to develop high performance timing protocol, also with respect to producing accelerator references with it.

## ACKNOWLEDGEMENTS

The presented work was made possible by the collaboration/ open source from ESS [20].